\documentstyle[11pt,newpasp,twoside,epsf]{article}
\def\edcomment#1{\iffalse\marginpar{\raggedright\sl#1\/}\else\relax\fi}
\reversemarginpar

\begin{document}
\title{Cosmic Distances: Current Odds and Future Perspectives}
 \author{Giuseppe Bono}
\affil{INAF - Rome Astronomical Observatory, Via Frascati 33, 
00040 Monte Porzio Catone, Italy; bono@mporzio.astro.it}

\begin{abstract}
We discuss recent theoretical and empirical results concerning cosmic 
distances and discuss the problems affecting the Cepheid and the RR Lyrae 
distance scales. In particular we outline the key role that first overtone 
Cepheids can play to improve the accuracy of distance determinations to  
nearby galaxies. We also address the positive features of the K-band 
Period-Luminosity-Metallicity ($PLZ_K$) relation of RR Lyrae stars 
when compared to the $M_V$ vs [Fe/H] relation. Moreover, we 
discuss the impact that accurate multiband data in external galaxies 
can have on the evolutionary properties of intermediate and low-mass 
stars. Finally, we introduce possible avenues of future research in 
stellar astrophysics.   
\end{abstract}

\section{Introduction}

Dating back to Baade (1944, 1958) the interplay between investigations 
on stellar populations and on radial variables as primary distance 
indicators is a well-established route to improve the knowledge 
on the stellar content of simple and complex stellar systems.  
During the last half century paramount observational and theoretical 
efforts have been devoted to single out the evolutionary and 
pulsation properties of population I and population II stars. 
In particular, spectroscopic measurements disclosed that the 
metallicity was not the leading physical parameter to 
split Baade's populations. Actually, Galactic RR 
Lyrae stars present a metallicity distribution that ranges from 
$[Fe/H]\approx-2.2$ up to solar metallicities (Preston 1964; 
Lub 1977). At the same time, Classical Cepheids have been discovered 
in dwarf galaxies such as Phoenix and IC~1613 that are 1 dex 
more metal-poor ($[Fe/H]\approx-1.9$; $[Fe/H]\approx-1.3$) than the 
Large and the Small Magellanic Cloud (LMC, $[Fe/H]\approx-0.3$; SMC, 
$[Fe/H]\approx-0.7$). This is the reason why 
we subdived the two most popular groups of variables located in the 
Cepheid instability strip in low and intermediate-mass radial variables.  

The microlensing experiments (EROS, OGLE, MACHO) provided an unprecedented 
amount of photometric 
data. The number of variables for which pulsation parameters 
(periods, mean magnitude and colors, amplitudes) are available have 
doubled and in some cases increased by one order of magnitude.  
In this context, HST also played a crucial role, and indeed the two 
fundamental projects on stellar distances succeeded in the measurement 
of Cepheids in several spirals of the Virgo cluster (Freedman et al. 2001; 
Saha et al. 2001). This unique opportunity allowed a substantial 
improvement in the calibration of secondary distance indicators, 
and in the evaluation of the Hubble constant.  

This is the coarse-grained scenario and no doubts that the perspectives 
for the near future are quite promising. However, the fine-grained 
scenario still presents several unsettled problems. In the following we 
discuss the impact that they might have on cosmic distance determinations 
and on stellar populations.  
 
\section{Classical Cepheids and intermediate-mass stars}

Classical Cepheids are the most popular distance indicators, since they 
are bright stars and they can be easily identified in the Galaxy, in 
Local Group (LG) galaxies, and in spiral galaxies of the Virgo 
and Fornax clusters. Pros and cons 
of the Cepheid distance scale have been discussed countless in the 
recent literature (Sandage et al. 1999; Tanvir 1999; Feast 1999; 
Freedman et al. 2001).   
Plain physical arguments suggest that individual Cepheid distances 
can be obtained using the Period-Luminosity-Color (PLC) relation. 
The use of the PLC relation is hampered by two problems: {\em i)} 
two accurate mean magnitudes are required; {\em ii)} the use of optical 
bands does require an accurate knowledge of the reddening. To overcome 
the first problem the two $H_0$ projects decided to properly sample 
the V band and then to adopt a constant amplitude ratio to improve 
the accuracy of the mean I magnitude. This empirical trick is 
supported by theoretical predictions. Two different routes can be 
followed to overcome the latter problem: {\em i)} near-infrared 
(NIR) magnitudes are marginally affected by reddening. However, 
we still lack accurate mean J and K magnitudes for Cepheids in 
external galaxies. Ground-based NIR observations with 8m class 
telescopes and/or H-band NICMOS observations will become available 
in the near future, but certainly NGST will play a major role to 
improve both the accuracy and the sample size.  
{\em ii)} the Wesenheit magnitudes (W=V-2.45(V-I)) are reddening 
free. This approach mimics the use of a PLC relation in the V, V-I 
band, but the color coefficient attain slightly different values.   
The strengths and weaknesses mentioned before are the main reasons why 
Cepheid distances are estimated using the Period-Luminosity 
(PL) relation. The main problem in using the PL relation is its intrinsic 
width in effective temperature. This thorny problem is severe in the 
optical but negligible in the NIR bands (Madore et al. 1987).  

On top of these problem there is the long-standing debate concerning 
the universality of both PL and PLC relations, i.e. the fact that the 
zero-point and the slopes of these relations might depend on the metal 
content. During the last few years this crucial point has been 
addressed in several theoretical and empirical investigations
(Sasselov et al. 1997; Kennicutt et al. 1998; Macri et al. 2001). These 
investigations do not agree on the magnitude of the effect, but they 
agree on the sign. In particular, they found that metal-rich Cepheids, 
at fixed period, are {\em brighter} than metal-poor ones.   
Theoretical models constructed by adopting different physical 
assumptions concerning the coupling between radial 
displacements and convection (Chiosi, Wood, \& Capitanio 1993; 
Bono et al. 1999a,b; Alibert et al. 1999) also do agree on the sign 
among themselves but do not agree on the magnitude of the metallicity 
dependence. What is really {\em puzzling} is that both linear and 
nonlinear Cepheid models are at odds concerning the sign, since 
they predict that metal-rich Cepheids, at fixed period, are 
{\em fainter} than metal-poor ones.  
Two pieces of evidence somehow support theoretical predictions: 
{\em i)} theory and observations suggest that metal-rich Cepheids 
are redder than metal-poor ones; {\em ii)} recent detailed analysis 
of V, I, K-band data of a sizable sample of MC Cepheids 
support the theoretical sign (Groenewegen \& Oudmaijer 2000;  
Groenewegen 2000; Storm et al. 2000). This problem requires a firm 
solution, since it affects the calibration of both zero-point and 
slope. The approach adopted by Pietrzynski et al. (2002) to measure 
Cepheids on a substantial portion of NGC~300 is quite promising to 
settle this problem.  

\begin{figure}  
\plotfiddle{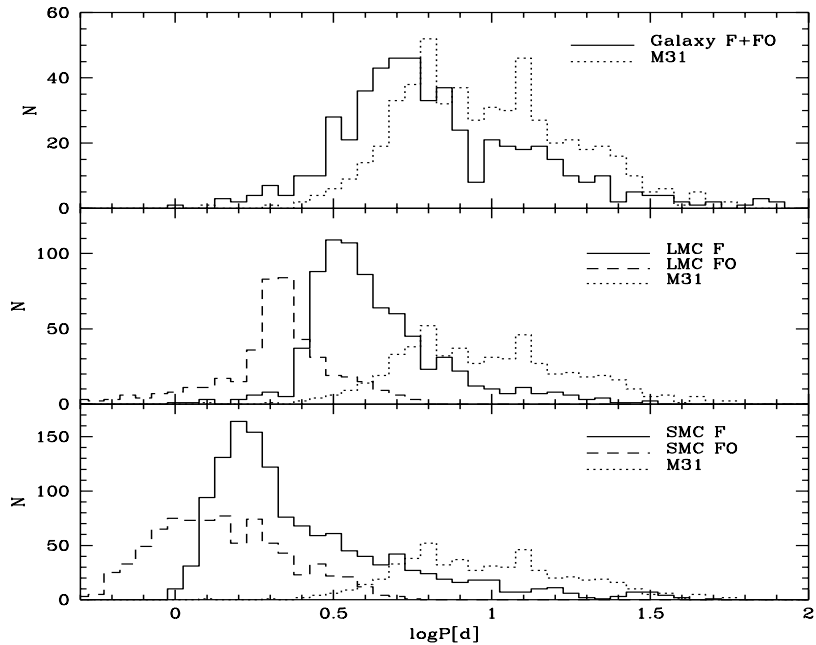}{0.5truecm}{0.}{57.}{46.}{-180.}{-290.}  
\vspace*{7.33truecm}\\  
\caption{  
\footnotesize 
Comparison between the period distribution of Cepheids in M31 
with Galactic (top) and Magellanic (middle and bottom) Cepheids.
Solid and dashed lines show the period distribution of fundamental and
first overtone respectively. The period distribution of Galactic Cepheids
accounts for both F and FO Cepheids. It is not clear whether the current
sample of M31 Cepheids (Baade \& Swope 1965; Mochejska et al. 2001) 
includes FOs (dotted line).}  
\end{figure}

However, recent theoretical and empirical results do suggest that 
the metal content marginally affects the PL and the PLC relation 
of First Overtone (FO) Cepheids. As a matter of fact, Bono et al. 
(2002) found that predicted and empirical $PL_K$ relations (K band 
data from the Two Micron All Sky Survey) and Wesenheit functions 
(V,I band data from the Optical Gravitational Lensing Experiment)  
supply mean distances to the Magellanic Clouds (MCs) that agree very 
well with each other. In particular they estimated a distance for 
the LMC  of $18.53\pm0.08$ (theory) and $18.48\pm0.13$ (obser.), 
as well as for the SMC of $19.04\pm0.11$ (theory) and $19.01\pm0.13$ 
(obser.).   The reasons for the agreement are manifolds: 
{\em i)} K-band data and Wesenheit function are marginally affected 
by uncertainties on reddening corrections;
{\em ii)} The $PL_K$ relation and the Wesenheit function presents 
a mild dependence on metal content (Bono et al. 1999b).   
{\em iii)} The previous intrinsic features do apply to fundamental (F) 
Cepheids, however, they are magnified for FO Cepheids, since the width 
in temperature of the latter pulsators is systematically smaller than 
for the former ones. Therefore, distances based on these variables 
are less affected by the typical spread of PL relations.  

The improvement in using FOs to constrain the accuracy of the Cepheid 
distance scale is undoubtful. The simultaneous 
detection of both F and FO Cepheids in nearby stellar systems should 
allow us to estimate on a quantitative basis the systematic errors 
affecting optical and NIR PL/PLC relations. Unfortunately, current 
photometric data for FO Cepheids are scanty, and indeed we still lack 
detailed information for FOs in nearby dwarfs such as Phoenix 
(Caldwell et al. 1988) but also for spiral galaxies such as M~31. 
Fig. 1 shows 
the period distribution of F and FO Cepheids in the Galaxy, M31, and 
in the MCs. Data plotted in this figure clearly show that the mean 
metallicity affects the period distribution. In fact, the peak 
decreases from $\log P=0.7-0.8$ in the Galaxy and in M~31 
($Z\approx0.02$) to $\log P=0.2$ in the SMC ($Z\approx0.004$). 
The same outcome applies to FO periods, and indeed the period 
distribution peaks around $\log P=0.3$ (LMC) and around 
$\log P=0.0-0.1$ (SMC). The sample of Galactic FOs is too 
small to draw any firm conclusion. The main drawback that limits 
the use of FOs is that they present short periods (see Fig. 1), 
and, therefore, they are on average 1-2 mag fainter than F variables 
(see Fig. 2). 

\begin{figure}  
\plotfiddle{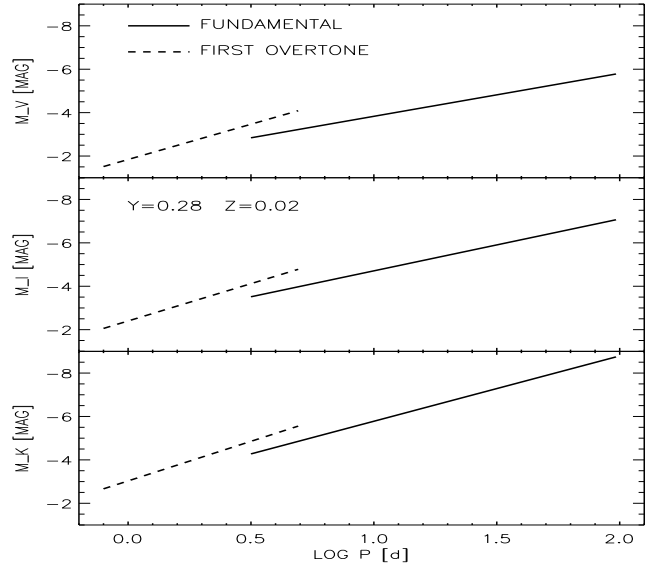}{0.5truecm}{0.}{50.}{33.}{-140.}{-210.}  
\vspace*{6.40truecm}\\  
\caption{
\footnotesize 
Theoretical PL relations in three different photometric 
bands V (top), I (middle), and K (bottom) constructed by adopting 
a solar chemical composition and a canonical ML relation (Bono et 
al. 2002). Solid and dashed lines display the F and FO PL relations.} 
\end{figure}

The detection of a sizable sample of FOs in nearby galaxies is important  
to improve the accuracy of Cepheid distances as well as  
to improve our knowledge on He-burning phases and on the occurrence of 
blue loops. The lower limit in the period distribution of FOs depends 
on the minimum 
mass whose blue loop crosses the instability strip. Current evolutionary 
predictions (Bertelli, Bressan, \& Chiosi 1985; Castellani, Chieffi, \& 
Straniero 1990; Stothers \& Chin 1993; Bono et al. 2000; Limongi et al. 
2000) suggest that the occurrence and the extent in temperature of blue 
loops strongly depend on the chemical compositions, on the efficiency of 
the mass loss as well as on the input physics and on the physical 
assumptions adopted to handle mixing processes in the convective layers.  

Up to now a detailed comparison between theory and observations has been 
performed for field stars in the MCs and for a 
few LMC clusters and in particular for NGC~1866. Although several  
theoretical and observational (HST) studies have been devoted to this 
cluster we still lack firm quantitative constraints on the size of the 
convective core around the Turn-Off region (Testa et al. 1999; Barmina 
et al. 2002; Walker et al. 2002).  
This cluster might be crucial to address some problems 
concerning the Main Sequence fitting and the Cepheid distance 
scale, since it hosts more than 20 Cepheids (Walker et al. 2001). However, 
the pulsation parameters are only available for half of them and we still 
lack homogeneous photometric data for bright and faint cluster stars.  

\section{RR Lyrae variables and low-mass stars}

RR Lyrae stars are very good tracers of the old stellar component, and 
play a key role in the absolute age determination of Globular Clusters. 
Although, RR Lyrae stars present several undoubtful advantages, the 
difference between distance estimates based on different calibrations 
(Baade-Wesselink method, HB models) of the 
$M_V$ vs [Fe/H] relation is systematically larger than the empirical 
uncertainties, and cover a range of $\approx 0.3$ mag on the distance 
modulus. This indicates that either the reddening accuracy is poor 
or RR Lyrae distances are still affected by systematic uncertainties. 
Moreover, recent theoretical (Caputo et al. 2000) and empirical
(Layden 2000) studies confirm that the $M_V$ vs [Fe/H] relation
is not linear when moving from metal-poor to metal-rich RR Lyrae. 

Some of the problems affecting the RR Lyrae distance scale can be
overcome using the K-band Period-Luminosity ($PL_K$) relation. In a 
seminal empirical investigation Longmore et al. (1990) demonstrate 
that RR Lyrae do obey a $PL_K$ relation in this band.
This finding was further strengthened by a recent theoretical
investigation (Bono et al. 2001) suggesting that RR Lyrae obey to 
a very tight $PLZ_K$ relation connecting the period, the luminosity, 
the K-band absolute magnitude, and the metallicity. Current models also 
disclosed that the $PLZ_K$ relation is, in contrast with the 
$M_V$ vs [Fe/H] relation, marginally affected by off-ZAHB evolution 
and by a spread in stellar masses (see Fig. 3).   
The intrinsic accuracy of previous predictions was tested using the 
prototype RR Lyr. Recently  Benedict et al. (2002) on the basis of 
new astrometric data collected with FGS~3@HST, provided an accurate 
estimate of the absolute trigonometric parallax of RR Lyr. The 
uncertainty of the new parallax ($\pi_{abs}=3.82\pm0.20$ mas) 
is $\approx 3$ times smaller than the uncertainty on 
the Hipparcos parallax ($\pi_{abs}=4.38\pm0.59$ mas). 
Interestingly, Bono et al. (2002) using the predicted 
$PLZ_K$ relation found a pulsation parallax $\pi_{puls}=3.858+/-0.131$ mas 
which agrees quite well with the HST parallax and presents a smaller 
formal error.  
 
\begin{figure}  
\plotfiddle{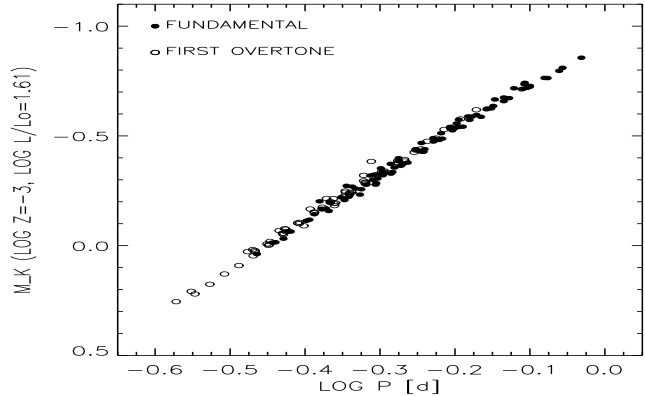}{0.5truecm}{0.}{50.}{27.}{-150.}{-160.}  
\vspace*{4.33truecm}\\  
\caption{
\footnotesize 
Projection onto a bidimensional plane of the predicted $PLZ_K$ 
relation for F and FO pulsators. $M_K$ 
magnitudes were scaled to the same metallicity 
($\log Z=-3$), and stellar luminosity ($\log L/L_\odot=1.61$). FO periods 
were  fundamentalized ($\log P_F = \log P_{FO} + 0.127$). Pulsation models 
cover a wide range in metallicity ($0.0001 \le Z \le 0.02$), stellar mass
($0.53 \le M/M_\odot \le 0.75$), and luminosity.} 
\end{figure}

It is noteworthy that HB stars are the crossroad of several astrophysical 
problems such as the second-parameter problem, the UV-upturn in ellipticals 
(Ferguson, this volume), as well as the dependence of $M_V(RR)$ on 
metallicity. Detailed extragalactic 
samples of HB and RR Lyrae stars are only available for MCs and a few 
LG dwarfs (Mateo 1998; Bersier \& Wood 2002). New homogeneous data 
of HB stars in stellar systems with different chemical and dynamical 
histories might play a crucial role to constrain their evolution properties.  

\section{Final remarks}

Some problems affecting the evolutionary and pulsation properties of 
classical Cepheids and RR Lyrae stars can be settled during the next 
ten years using the unprecedented sensitivity and spatial resolution 
of the CCD camera that are available (ACS) or will become available 
(WFPC3) on board of HST. Sizable samples of FO Cepheids might be 
detected and measured across the disk of M~31. The number of orbits  
necessary to accomplish this experiment is rather modest, since this 
galaxy hosts regions rich of Cepheids and also because the 
apparent magnitudes around the minimum luminosity, assuming a distance 
modulus of 24.5 and a mean reddening of E(B-V)=0.08, roughly range from 
$V\approx22$ ($\log P =0.6$) to $V\approx24$ ($\log P =-0.1$).   
The same conclusions apply to RR Lyrae stars, and indeed their apparent 
magnitudes range from $V\approx25$ to $V\approx26$ (Clementini et al. 
2001). Note that up to now we still lack complete  sample of HB and 
RR Lyrae stars in elliptical galaxies such as M~32.  

A substantial improvement concerning the intrinsic accuracy of the 
Cepheid distance scale will also be provided by NGST. The superior 
sensitivity in the NIR bands will allow us to 
provide accurate measurements of Cepheids mean K magnitudes up to 
the Virgo and the Fornax cluster. The new data will supply robust 
distance determinations based on both the $PL_K$ and the $PLC (V,V-K)$  
relations. At the same time, NGST will supply the unprecedent opportunity 
to estimate the distance of several galaxies up to the outskirts of the 
LG using simultaneously NIR relations for RR Lyrae ($PLZ_K$) and 
Classical Cepheids.  
Future astrometric missions such as the SIM, and GAIA may 
also supply accurate empirical constraints concerning the accuracy of 
absolute (zero-point) and relative (slope) distance estimates  
as well as on their dependence on metallicity.  

The previous improvements would supply tight constraints on the systematic 
uncertainties affecting primary distance indicators, and hopefully a 
determination of the Hubble constant with a global accuracy of the order 
of 5-7\%. The next step that can allow us to nail down the systematic 
uncertainty affecting the evaluation of $H_0$ is to only use Cepheids 
and to by pass secondary distance indicators. Data available in the 
literature  clearly show that distance determinations to the Coma cluster 
based on different zero-points and secondary indicators range from 
$34.64\pm0.25$ (SBF in the K-band, Jensen et al. 1999) to $35.29\pm0.11$ 
(various methods, Tammann et al. 1999), while the $H_0$ values range 
from $85\pm10$ to $60\pm6$ $km\,s^{-1} Mpc^{-1}$.

The main advantage in using Classical Cepheids is that we know the physical 
mechanisms that drive the pulsation instability in these objects. On the 
other hand, the experiment is challenging from an observational point of 
view, since it is necessary to detect and measure Cepheids 
in giant spirals of the Coma cluster. This means that we should be able to 
perform accurate photometry down to $V\approx31-32$ mag with a spatial 
resolution that is at least a factor of five better than current HST 
capabilities. We also note that this limit magnitude would supply 
the unique opportunity to estimate the age of old stellar component in 
LG galaxies up to the outer fringes (NGC~3109, Antlia), and in turn 
to supply a robust lower limit to the age of the Universe. Even though 
we expect that these two parameters are tightly correlated we could 
agree with Baade's statement: {\em Although I am quite certain that 
I do not mistake pink elephants for pink mice an unprejudiced check 
is always reassuring} (as quoted by Feast 2000).  
 
It is a pleasure to thank N. Panagia for a detailed reading of a  
draft of this paper. I am indebted to my collaborators for helpful 
discussions and suggestions.  

\end{document}